\documentclass[11pt,final,journal,twoside]{IEEEtran}
\usepackage{blindtext}
\usepackage{titling}
\usepackage{array}
\usepackage{tabularx}
\usepackage{booktabs}
\usepackage{pifont}
\usepackage{hyperref}
\preauthor{\begin{center}
    \large \lineskip .75em%
    \begin{tabular}[t]{>{\centering\arraybackslash}p{.45\textwidth}}}
\postauthor{\end{tabular}\par\end{center}}

\usepackage[table,html]{xcolor}
\usepackage[most]{tcolorbox}
\definecolor{viridis1}{HTML}{440154} 
\definecolor{viridis2}{HTML}{3B528B} 
\definecolor{viridis3}{HTML}{21918C}
\definecolor{viridis4}{HTML}{5DC863}
\definecolor{viridis5}{HTML}{FDE725} 

\makeatletter
\renewcommand\and{
  \end{tabular}%
  \hfill
  \begin{tabular}[t]{>{\centering\arraybackslash}p{.45\textwidth}}}
  \makeatother
\usepackage{amsfonts,amsmath,amssymb,amsthm} 
\usepackage{enumitem}
\usepackage{latexsym} 
\usepackage{mathrsfs}
\usepackage{graphicx}
\usepackage{indentfirst}
\usepackage{fullpage}
\usepackage{url}
\usepackage{multicol}
\usepackage{multirow}
\usepackage{color}
\usepackage{tikz} 
\usepackage{pgfplots}
\usetikzlibrary{arrows.meta}
\usepackage{cleveref}
\usepackage{siunitx}
\usetikzlibrary{positioning, fit, backgrounds, calc}
\usepackage{tikz-3dplot}
\usepackage{caption}
\usetikzlibrary{arrows.meta, positioning, shapes.geometric, calc}

\newcolumntype{Y}{>{\raggedright\arraybackslash}X}
\newcolumntype{L}{>{\raggedright\arraybackslash}p{0.22\textwidth}}
\newcolumntype{M}{>{\raggedright\arraybackslash}p{0.36\textwidth}}
\newcolumntype{R}{>{\raggedright\arraybackslash}p{0.3425\textwidth}}

\usepackage[
  n,
  advantage,
  operators,
  sets,
  adversary,
  landau,
  probability,
  notions,
  logic,
  ff,
  mm,
  primitives,
  events,
  complexity,
  asymptotics,
  keys
]{cryptocode}
\renewcommand{\verify}{\ensuremath{\mathsf{Verify}}}
\renewcommand{\hash}{\ensuremath{\mathsf{Hash}}}

\newcommand{\rmv}[1]{}

\renewcommand{\vec}[1]{\ensuremath{\overrightarrow{#1}}}

\newcommand{\F}{\mathbb{F}}

\newcommand{\x}{\mathbf{x}}
\newcommand{\s}{\mathbf{s}}
\newcommand{\w}{\mathbf{w}}
\newcommand{\y}{\mathbf{y}}

\newcommand{\e}{\ensuremath{\mathbf{e}}}

\newcommand{\rref}{\ensuremath{\mathsf{rref}}}
\newcommand{\Aut}{\ensuremath{\mathrm{Aut}}}

\newcommand{\lang}{\ensuremath{\mathcal{L}}}
\newcommand{\relation}{\ensuremath{\mathcal{R}}}

\newcommand{\rsends}[2][0.2\textwidth]{\ensuremath{\xrightarrow[\hspace{#1}]{ #2 }}}
\newcommand{\lsends}[2][0.2\textwidth]{\ensuremath{\xleftarrow[\hspace{#1}]{ #2}}}

\newcommand{\chal}{\ensuremath{\mathsf{chal}}}
\newcommand{\comm}{\ensuremath{\mathsf{comm}}}
\newcommand{\resp}{\ensuremath{\mathsf{resp}}}

\newcommand{\rank}{\ensuremath{\mathrm{rank}}}

\newcommand{\wt}{\ensuremath{{wt}}}
\newcommand{\Mon}[1][n]{\ensuremath{\mathrm{Mon}_{#1}}}
\newcommand{\GL}{\ensuremath{\mathrm{GL}}}

\renewcommand{\Vec}{\ensuremath{\mathrm{vec}}}

\newcommand{\Ext}{\ensuremath{\mathrm{Ext}}}
\newcommand{\Sim}{\ensuremath{\mathrm{Sim}}}

\theoremstyle{remark}

\title{Digital signature schemes based on code equivalence and syndrome decoding from restricted errors
\thanks{
The authors are with the Department of Mathematics, Virginia Tech (email: \texttt{\{sarpin, jlegrow, hhlopez, gmatthews\}@vt.edu}). 
The National Science Foundation partially supported the third  (DMS-2401558 and 2502705) and fourth (DMS-2201075 and 2502705) authors. All four authors are partially supported by the Commonwealth Cyber Initiative.
}}

\author{Sarah Arpin \and Jason T.~LeGrow \and Hiram H. L\'opez \and Gretchen L. Matthews}

\date{}

\begin{document}

\maketitle

\begin{abstract}
\rmv{This article focuses on how error-correcting codes may be used to design digital signature schemes. Because} Digital signature schemes are an important cryptographic tool to ensure data authenticity and integrity in many applications that must be resilient to attacks, including those facilitated by quantum computers. We consider the two digital signature schemes based on error-correcting codes that are second-round candidates in NIST's call for Additional Signature Schemes, which is part of the Post-Quantum Cryptography Standardization Process. Specifically, we provide an overview of the Codes and Restricted Objects Signature Scheme (CROSS) and the Linear Equivalence Signature Scheme (LESS). We describe their underlying problems of syndrome decoding from restricted errors and code equivalence. We review sigma protocols and how they can be transformed into digital signature schemes via the Fiat-Shamir transform. Finally, we explain how this procedure yields code-based digital signatures believed to be post-quantum secure. 
\end{abstract}

\section{Introduction} \label{section:intro}
This article focuses on digital signature schemes based on error-correcting codes. Digital signatures are much like handwritten signatures but designed for technology. They support authentication, ensuring data origin; integrity, confirming that data has not been tampered with or modified; and accountability, also called non-repudiation, so that parties cannot later deny having sent or signed a message. Digital signatures underpin applications from the internet to blockchain (Table~\ref{tab:dss_uses}). Given recent advances in quantum computing, which include  period‑finding algorithms (e.g., Shor's Algorithm \cite{Shor_94}) and search algorithms that demonstrate quantum speedup (e.g., Grover’s algorithm \cite{Grover}), digital signatures must remain secure even against attacks enabled by the most modern or emerging technologies. In this work, we discuss recent developments in using error-correcting codes to design digital signature schemes believed to be quantum-safe, meaning they are resilient against attacks from both classical and quantum algorithms. In particular, we focus on the two code-based digital signature schemes being considered by the National Institute of Standards and Technology (NIST) in the second round of Additional Signature Schemes \cite{NIST_PQC_DSS_Round1_report}, which is part of the Post-Quantum Cryptography Standardization Process \cite{nist_pqc_standardization}.
\rmv{
\begin{table*}[t]
  \centering
  \begin{tabularx}{\textwidth}{L M  R }
\textbf{Domain} & \textbf{Common Uses \& Examples } & \textbf{Purpose} \\ \hline
\midrule
Secure Communication &
email (S/MIME), secure messaging, digitally signed documents &
 certificates, message integrity, and signer authentication\\ \hline
Software \& Content &
code signing, package signing, firmware validation &
 package manager signatures, verify publisher and integrity \\ \hline
Finance \& Transactions &
e-banking approvals, e-commerce payment authorization, transaction receipts &
non-repudiation of approvals, sometimes combined with hardware tokens \\ \hline
Legal \& Government &
contracts and agreements, tax filing, digital notarization &
compliance frameworks,  advanced or qualified signatures, audit trails\\ \hline
Web Security &
TLS/HTTPS certificates, server/client authentication, API trust &PKI, certificate chains,  signatures bind public keys to identities \\ \hline
Documents &
 PDF signing, time stamping authority, long-term validation &time-stamping, embedded revocation data \\ \hline
Blockchain &
transaction signing, wallets, smart contracts authorization & signatures authorize on-chain state changes \\ \hline
IoT / Embedded Systems &
secure boot, signed firmware/ updates, device attestation &
verify signatures prior to execution, supply-chain integrity, key storage \\ \hline \\
  \end{tabularx}
  \caption{Common Uses and Purposes of Digital Signatures}
  \label{tab:dss_uses}
\end{table*}
}
\renewcommand{\arraystretch}{1.2}
\begin{table*}[t]
\centering
\rowcolors{2}{gray!10}{white}
\begin{tabularx}{\textwidth}{L M R}
\textbf{Domain} &
\textbf{Common Uses \& Examples} &
\textbf{Purpose} \\

\toprule

Blockchain &
Transaction signing, wallets, and smart contract authorization. &
Authorization of on-chain state changes. \\

Documents &
PDF signing, time-stamping authorities, and long-term validation. &
Trusted timestamps and embedded revocation data. \\

Finance \& Transactions &
E-banking approvals, e-commerce payments, and transaction receipts. &
Non-repudiation and hardware tokens. \\

IoT / Embedded Systems &
Secure boot, signed firmware updates, and device attestation. &
Execution integrity, supply-chain security, and key protection.\\

Legal \& Government &
Contracts and agreements, tax filing, and digital notarization. &
Regulatory compliance frameworks, qualified signatures, and audit trails. \\

Secure Communication &
Email (S/MIME), secure messaging, and digitally signed documents. &
Certificates, message integrity, and signer authentication. \\

Software \& Content &
Code signing, package signing, and firmware validation. &
Package manager signatures, verify publisher identity, and content integrity. \\

Web Security &
TLS/HTTPS certificates, server/client authentication, and API trust. &
PKI, certificate chains, and binding public keys to identities. \\

\bottomrule
\end{tabularx}
\captionsetup{skip=0.3cm}
\caption{Common uses and purposes of digital signatures.}
\label{tab:dss_uses}
\end{table*}

Coding theory and cryptography evolved largely in parallel, with error-correcting codes enabling reliable transmission over noisy channels and cryptosystems protecting data from unauthorized parties. Their intersection emerged with the public-key proposals of McEliece~\cite{McEliece} and Niederreiter~\cite{NIEDERREITER_86}, which exploit noise, code structure, and decoding to achieve secure communication. Despite this promise, code-based schemes were not incorporated into international standards, largely due to their large key sizes and the availability of more efficient alternatives such as RSA~\cite{RSA}, elliptic curve cryptography, and Diffie–Hellman~\cite{diffie_hellman_1976}. Consequently, 
code-based schemes were long viewed as theoretically appealing but less practical than competing implementations.

Post-quantum (also called quantum-safe or quantum-resistant) cryptosystems are designed to withstand attacks by both classical and quantum algorithms. In 1994, Shor introduced a quantum algorithm~\cite{Shor_94} that runs in polynomial time for integer factorization and for discrete logarithms (including the elliptic-curve variant), implying that widely deployed schemes such as RSA, Diffie–Hellman, and elliptic curve cryptography are not post-quantum. These results motivate new public-key primitives and have renewed interest in code-based constructions, where error-correcting codes remain among the most promising tools for post-quantum cryptography.

Much of code-based cryptography targets key encapsulation mechanisms (KEMs) that let two parties establish a shared secret despite eavesdroppers, enabling  secure communication. Several proposals—including Classic McEliece \cite{classic_mceliece_spec} and BIKE \cite{bike_spec}—follow the McEliece/Niederreiter paradigm of hiding a structured, efficiently decodable code, while HQC \cite{hqc} instead relies on syndrome decoding and was selected by NIST for standardization.

\newcommand{\schemename}[1]{{\bfseries\normalsize #1}}
\newcommand{\hardness}[1]{{\scriptsize\itshape #1}}
\begin{figure}
  \centering
  \resizebox{\columnwidth}{!}{%
  \begin{tikzpicture}[
    font=\footnotesize,
    title/.style={font=\bfseries\normalsize},
    subtitle/.style={font=\footnotesize\bfseries},
    hardbox/.style args={#1}{draw, thick, rounded corners, fill=#1,
      minimum width=4.0cm, minimum height=1.45cm,
      inner sep=6pt, align=center}
  ]
 
  \node[title] (classical_title) {Classical Digital Signatures};
  \node[subtitle, below=1pt of classical_title] (classical_sub)
    {\it mathematics underpinning security};

  \node[title, right=2cm of classical_title] (pq_title)
    {Post-Quantum Digital Signatures};
  \node[subtitle, below=1pt of pq_title] (pq_sub)
    {\it mathematics underpinning security};

  \node[hardbox={red!12}, below=0.55cm of classical_sub] (rsa) {
    \schemename{RSA}\\[2pt]
    \hardness{\it Integer Factorization}
  };

  \node[hardbox={red!12}, below=0.55cm of rsa] (ecdsa) {
    \schemename{DSA}\\[2pt]
    \hardness{\it Discrete Logarithm Problem}
  };

  \node[hardbox={red!12}, below=0.55cm of ecdsa] (eddsa) {
    \schemename{ECDSA/EdDSA}\\[2pt]
    \hardness{\it Elliptic Curve Discrete Logarithm Problem}
  };

  \node[hardbox={green!15}, below=0.55cm of pq_sub] (dilithium) {
    \schemename{Dilithium, Falcon}\\[2pt]
    \hardness{\it Module Learning with Errors / Lattices}
  };

  \node[hardbox={green!15}, below=0.55cm of dilithium] (falcon) {
    \schemename{Unbalanced Oil and}\\[2pt]
    \schemename{Vinegar (UOV) }\\[2pt]
    \hardness{\it Multivariate Polynomial Systems}
  };

  \node[hardbox={green!15}, below=0.55cm of falcon] (sphincs) {
    \schemename{CROSS}\\[2pt]
    \hardness{\it Restricted Syndrome Decoding}
  };

  \node[hardbox={green!15}, below=0.55cm of sphincs] (wavecross) {
    \schemename{LESS, MEDS}\\[2pt]
    \hardness{\it Code Equivalence}
  };

  \begin{pgfonlayer}{background}
    \node[
      draw, thick, rounded corners, fill=red!6,
      fit=(classical_title)(classical_sub)(rsa)(eddsa),
      label=below:
    ] {};
    \node[
      draw, thick, rounded corners, fill=green!8,
      fit=(pq_title)(pq_sub)(dilithium)(wavecross),
      label=below:
    ] {};
  \end{pgfonlayer}
  \end{tikzpicture}
 }
  \caption{Examples of digital signature schemes. \rmv{grouped by security regime}}
  \label{fig:selected_dss_sec}
\end{figure}
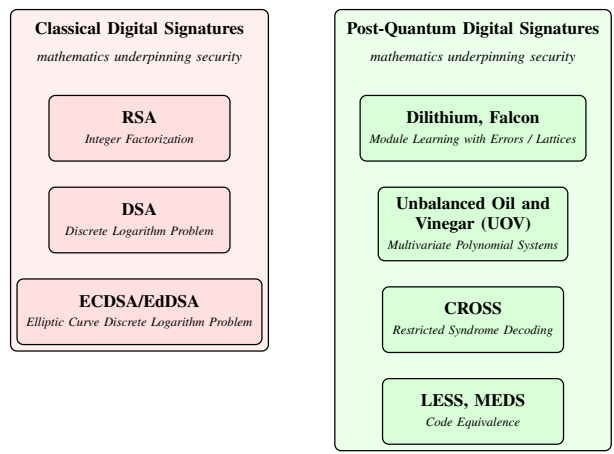

The motivation for this expository paper is multifold. First, there is an urgency to transition to quantum-safe cryptography and infrastructure, given new requirements and recommendations such as those from the White House and the European Commission. The National Institute of Standards and Technology (NIST) recommends transitioning by 2030 and requires it by 2035~\cite{moody2024nist8547}. Most current digital signatures are created using classical protocols such as those shown in Figure~\ref{fig:selected_dss_sec}. Second,  digital signature schemes are ubiquitous, and our digitally connected world, computing systems, and critical infrastructure relies on them. 
They support secure network communications, authentication of software and firmware updates, and control command verification and data within mission‑critical cyber‑physical systems.
   Third, this topic opens new problem spaces in which individuals with interests in information theory and related topics may play a role. 
As most code-based cryptography literature focuses on public-key encapsulation mechanisms, we concentrate on code-based signature schemes. This article aims to fill a gap in the literature, providing a new point of entry for those curious about how error-correcting codes are used in digital signature schemes. 

We focus on code-based digital signatures, showcasing those that remain as candidates in Round 2 of  NIST's Standardization Process for Additional Digital Signatures, namely the Codes and Restricted Objects Signature Scheme (CROSS) \cite{cross_spec} and the Linear Equivalence Signature Scheme (LESS) \cite{AFRICACRYPT:BMPS20}. It is worth noting that NIST is specifically interested in signature schemes that are not based on structured lattices, likely because the lattice-based schemes CRYSTALS-Dilithium \cite{Dilithium} and FALCON \cite{Falcon} have been approved for standardization.

In the next section, we review digital signature schemes and zero-knowledge proofs followed by how they are used to provide digital signatures via the Fiat-Shamir transform. We survey the Restricted Syndrome Decoding Problem and the digital signature scheme CROSS. This is followed by an overview of the digital signature scheme LESS, which is based on the code equivalence problem. The article concludes with a brief summary.

\section{Digital Signature Schemes and Zero-knowledge Protocols}
\label{section:ds_zk}
Digital signatures are used primarily for:
\begin{itemize}
    \item \emph{Data Integrity}, ensuring that messages are not tampered with in transit between parties;
    \item \emph{Data Origin Authentication}, verifying the sender of a message; and,
    \item \emph{Non-Repudiation}, guaranteeing that a user cannot later deny having signed a message.
\end{itemize}
Formally, a~\emph{digital signature scheme} is a cryptographic protocol with the following components:
\begin{figure*}[t]
  \centering
  \resizebox{\textwidth}{!}{%
  \begin{tikzpicture}[
    entity/.style={draw, thick, rounded corners, minimum width=3.2cm, minimum height=1cm, align=center, fill=#1!15, font=\small},
    process/.style={draw, fill=#1!15, ellipse, minimum width=3cm, minimum height=1cm, align=center},
    data/.style={draw, fill=gray!20, rounded corners, minimum width=3cm, minimum height=0.8cm, align=center},
    adversary/.style={draw, thick, dashed, fill=red!10, rounded corners, minimum width=3.5cm, minimum height=1.5cm, align=center},
    arrow/.style={->, very thick, >=Latex},
    genuine/.style={arrow, green!50!black},
    forged/.style={arrow, red!70!black, dashed},
    font=\small
]

\node[entity=green] (alice) at (0,7.5) {Alice (signer)}; 
\node[process=green] (sign) at (0,5.5) {Sign with Private Key $sk_A$}; 
\node[adversary] (eve) at (0,3.5) {Eve (quantum-equipped adversary)}; 

\node[data] (msg) at (6,5.5) {$(m,\sigma)$};          
\node[data] (forge) at (6,3.5) {$(m',\sigma')$ (forged)}; 

\node[entity=cyan] (bob) at (12,7.5) {Bob (verifier)};   
\node[process=cyan] (verify) at (12,4.5) {Verify with Public Key $pk_A$}; 

\node[data, draw, fill=green!15, rounded corners, font=\scriptsize, inner sep=2pt] (accept) at (14,2.5) {Genuine $(m,\sigma)$};
\node[data, draw, fill=red!15, rounded corners, font=\scriptsize, inner sep=2pt] (reject) at (10,2.5) {Forged $(m',\sigma')$};

\draw[arrow, green!50!black] (alice.south) -- (sign.north);

\draw[genuine, bend left=10] (sign.east) -- (msg.west);

\draw[forged, bend right=10] (eve.east) -- (forge.west);

\draw[genuine, bend left=15] (msg.east) -- (verify.west);
\draw[forged, bend right=15] (forge.east) -- (verify.west);

\draw[arrow, cyan!70!black] (bob.south) -- (verify.north);

\draw[genuine] (verify.south) -- (accept.north) node[midway, right] {Accept};
\draw[forged] (verify.south) -- (reject.north) node[midway, left] {Reject};

\end{tikzpicture}%
  }
  \caption{Signing and verification with forgery  attempt; valid signatures verify and forgeries fail.}
  \label{fig:dss_e_forge}
\end{figure*}
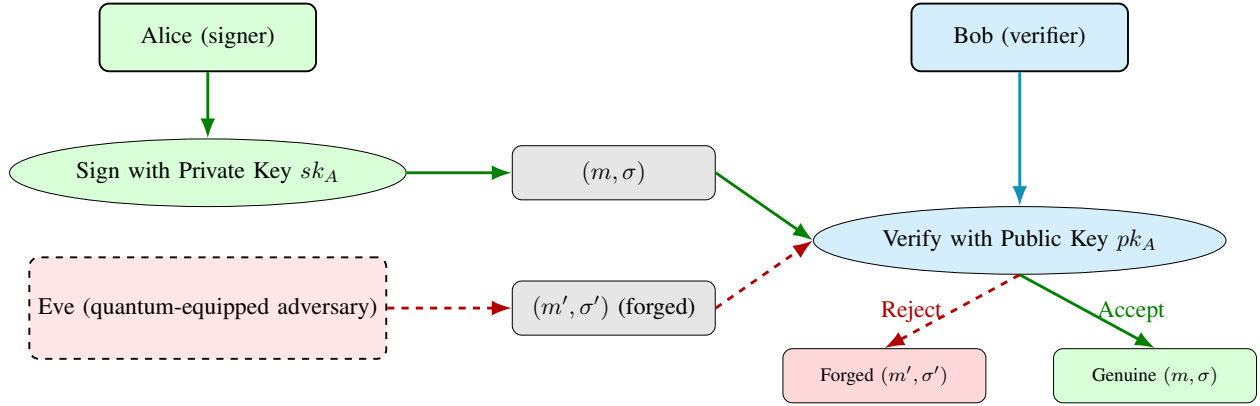
\begin{enumerate}[wide, leftmargin=0pt]
\item \textbf{Public parameter generation}: Given a security parameter \(\lambda\), outputs public parameters.
\item \textbf{Key generation}: Given public parameters, outputs a private (signing) key and a public (verification) key.
\item \textbf{Signing}: Given a secret key, corresponding public key, and a message, outputs a signature on the message.
\item \textbf{Verification}: Given a public key, a message, and a signature, either accepts or rejects the validity of the signature.
\end{enumerate}

The process and role of the public and private keys are illustrated in Figure \ref{fig:dss_e_forge}. Here, signer Alice holds the private (secret) key $sk_A$, which is associated with the public key $pk_A$; anyone can access the public key, while only Alice knows the private key. Alice signs message $m$ using her private key $sk_A$, resulting in signature $\sigma$. Then Bob (or any other verifier) may use the message-signature pair $(m,\sigma)$ and Alice's public key $pk_A$ to confirm the message was signed by Alice. The security here is based on the fact that an unauthorized party, such as Eve in Figure~\ref{fig:dss_e_forge}, cannot determine the private key from knowledge of the public key. Suppose that Eve has another message $m'$ that she wishes to pass off as sent (signed) by Alice. In this case, Eve shares $(m',\sigma')$ with Bob, who accepts that pair as sent by Alice if Eve is able to use $sk_A$ to sign it. Otherwise, Bob rejects the pair.

We require signature schemes to be \emph{correct}, meaning that if the signer follows the protocol honestly, then the verifier accepts the resulting signature. The typical security notion is \emph{existential unforgeability under chosen-message attack} (EUF-CMA), which intuitively states that an adversary who sees a number of valid message-signature pairs cannot produce a signature for a new message.

Zero-knowledge protocols are used to convince an entity that a statement is true without revealing any additional information. As an albeit overly  simplified example, consider a safe with an old-fashioned rotary (or dial) lock for which Alice knows the combination. Bob can challenge Alice by locking it and rotating the dial. She can demonstrate to Bob that she knows the combination by showing him the unlocked safe. Bob can repeat the challenge by rotating the dial to different locations, and equipped with the combination, Alice can unlock it. Each time, she shows Bob the unlocked safe. He is then convinced that Alice knows the lock combination, yet he has no knowledge of the combination itself.  Zero-knowledge protocols  employ mathematics to achieve this phenomenon of demonstrating knowledge without revealing the knowledge. Protocols should satisfy the properties found in Table~\ref{tab:zk_properties}. In Section~\ref{section:FS_transform}, we explain how zero-knowledge protocols can provide digital signature schemes.

\begin{table}[h]
  \centering
  \renewcommand{\arraystretch}{1.25}
  \rowcolors{2}{gray!10}{white}
  \begin{tabularx}{\columnwidth}{@{} l X @{}}
    \textbf{Property} & \textbf{Description} \\
    \toprule
    Completeness & \ding{51} Verifier accepts a true statement with high probability. \\
    Soundness & \ding{55} Dishonest prover cannot make the verifier accept a false statement, except with small probability. \\
    Zero-Knowledge & \ding{168} Verifier has no information beyond the veracity of the statement. \\
    \bottomrule
  \end{tabularx}
\caption{Zero-knowledge protocol properties, assuming honest prover.}
  \label{tab:zk_properties}
\end{table}

\section{Sigma Protocols and the Fiat-Shamir Transform} \label{section:FS_transform}

\begin{figure}[t]
\centering
\resizebox{\columnwidth}{!}{%
\rmv{
\begin{tikzpicture}[
  font=\footnotesize,
  party/.style={
    draw, thick, rounded corners,
    minimum width=1cm, minimum height=4.25cm,
    align=center
  },
  msg/.style={->, ultra thick, >=Latex},
  base/.style={->, very thick, >=Latex}
]

\node[party] (bob) {Bob\\\textit{(verifier)}};
\node[party, right=8.4cm of bob] (alice) {Alice\\\textit{(prover)}};

\coordinate (A1) at ([xshift=-3pt, yshift=45pt] alice.west);
\coordinate (B1) at ([xshift= 3pt, yshift=45pt]   bob.east);

\coordinate (B2) at ([xshift= 3pt, yshift=15pt]   bob.east);
\coordinate (A2) at ([xshift=-3pt, yshift=15pt] alice.west);

\coordinate (A3) at ([xshift=-3pt, yshift=-15pt] alice.west);
\coordinate (B3) at ([xshift= 3pt, yshift=-15pt]   bob.east);
\coordinate (BaseL) at ([xshift= 3pt, yshift=-45pt]   bob.east);
\coordinate (BaseR) at ([xshift=-3pt, yshift=-45pt] alice.west);

\draw[msg] (A1) -- node[above] {commitment $\comm$} (B1);
\draw[msg] (B2) -- node[above, sloped] {challenge $\chal$} (A2);
\draw[msg] (A3) -- node[above, sloped] {response $\resp_{\chal}$} (B3);


\draw[msg] (BaseL) -- node[above] {verification: accept/reject} (BaseR);

\end{tikzpicture}%
}
\begin{tikzpicture}[
    arrow/.style={->, ultra thick},
    party/.style={
        draw=gray!70,
        rounded corners=14pt,
        thick,
        fill=gray!20,
        minimum width=2cm,
        minimum height=3.5cm,
        align=center
    }
]

\node[party] (bob) at (0,0) {\Large{Bob}\\(verifier)};
\node[party] (alice) at (7,0) {\Large{Alice}\\(prover)};

\coordinate (commit_bob) at ($(bob.east)+(0,1.25)$);
\coordinate (commit_alice) at ($(alice.west)+(0,1.25)$);

\coordinate (challenge_bob) at ($(bob.east)+(0,0.85)$);
\coordinate (challenge_alice) at ($(alice.west)+(0,0.15)$);

\coordinate (response_bob) at ($(bob.east)+(0,-0.85)$);
\coordinate (response_alice) at ($(alice.west)+(0,-0.15)$);

\coordinate (verification_bob) at ($(bob.east)+(0,-1.25)$);
\coordinate (verification_alice) at ($(alice.west)+(0,-1.25)$);

\draw[arrow] (commit_alice) -- node[above] {Commitment $\comm$} (commit_bob);
\draw[arrow] (challenge_bob) -- node[yshift=0.2cm, sloped, rotate=-1] {Challenge $\chal$} (challenge_alice);
\draw[arrow] (response_alice) -- node[yshift=0.2cm, sloped, rotate=1] {Response $\resp_{\chal}$} (response_bob);
\draw[arrow] (verification_bob) -- node[below] {Verification: accept/reject} (verification_alice);

\end{tikzpicture}
}
\caption{Sigma‑protocol  depicting Alice using $sk_A$ to generate commitment $\comm$, followed by Bob sharing a challenge $\chal$, after which Alice provides a response $\resp_{\chal}$ using $sk_A$ that Bob can then verify using Alice's public key $pk_A$ leading to acceptance or rejection.} 
\label{fig:sigma_protocol}
\end{figure}
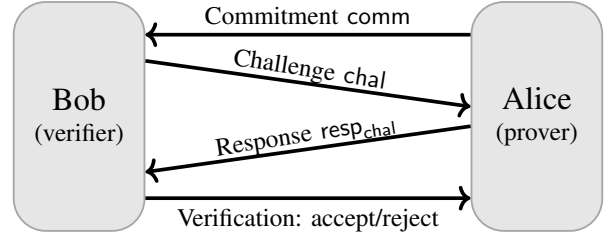

The Fiat-Shamir transform~\cite{Fiat_Shamir} turns interactive zero‑knowledge protocols into practical digital signature schemes; in particular, it converts a sigma protocol into a digital signature scheme. A \emph{sigma protocol} is an interactive protocol between two parties---the \emph{prover} \(\prover\) and the \emph{verifier} \(\verifier\)---in which the prover convinces the verifier that she knows a piece of secret information, without revealing the actual secret information. See Figure~\ref{fig:sigma_protocol} for a sketch. More than one-third of the submissions to NIST's Round 1 Additional Signatures incorporate the Fiat-Shamir transform, including CROSS and LESS, the code-based schemes discussed in this article. 

\begin{figure}[t]
\centering
\resizebox{\columnwidth}{!}{%
\rmv{
\begin{tikzpicture}[
  font=\footnotesize,
  party/.style={
    draw, thick, rounded corners,
    minimum width=1cm, minimum height=4.25cm,
    align=center
  },
  msg/.style={->, ultra thick, >=Latex},
  base/.style={->, very thick, >=Latex}
]

\node[party] (bob) {Bob\\\textit{(verifier)}};
\node[party, right=8.4cm of bob] (alice) {Alice\\\textit{(prover)}};

\coordinate (A1) at ([xshift=-3pt, yshift=30pt] alice.west);
\coordinate (B1) at ([xshift= 3pt, yshift=30pt]   bob.east);

\coordinate (B2) at ([xshift= 3pt, yshift=-24pt]   bob.east);
\coordinate (A2) at ([xshift=-3pt, yshift=-40pt] alice.west);

\coordinate (A3) at ([xshift=-3pt, yshift=-30pt] alice.west);
\coordinate (B3) at ([xshift= 3pt, yshift=-30pt]   bob.east);

\draw[msg] (A1) -- node[above] {$((\comm_1, \dots, \comm_{\kappa}), (\resp_{\chal_1}, \dots, \resp_{\chal_{\kappa}})) $} (B1);

\coordinate (BaseL) at ([xshift= 3pt, yshift=-86pt]   bob.east);
\coordinate (BaseR) at ([xshift=-3pt, yshift=-86pt] alice.west);

\draw[msg] (B3) -- node[above] {verification: accept/reject} (A3);

\end{tikzpicture}%
}
\begin{tikzpicture}[
    arrow/.style={->, ultra thick},
    party/.style={
        draw=gray!70,
        rounded corners=14pt,
        thick,
        fill=gray!20,
        minimum width=2cm,
        minimum height=3.5cm,
        align=center
    }
]

\node[party] (bob) at (0,0) {\Large{Bob}\\(verifier)};
\node[party] (alice) at (7,0) {\Large{Alice}\\(prover)};

\coordinate (commit_bob) at ($(bob.east)+(0,0.75)$);
\coordinate (commit_alice) at ($(alice.west)+(0,0.75)$);

\coordinate (verification_bob) at ($(bob.east)+(0,-0.75)$);
\coordinate (verification_alice) at ($(alice.west)+(0,-0.75)$);

\draw[arrow] (commit_alice) -- node[yshift=.75cm] {Commitment} (commit_bob);
\draw[arrow] (commit_alice) -- node[above] {\small{$(\comm_1, \dots, \comm_{\kappa}) $}} (commit_bob);
\draw[arrow] (commit_alice) -- node[below] {\small{$ (\resp_{\chal_1}, \dots, \resp_{\chal_{\kappa}}) $}} (commit_bob);
\draw[arrow] (verification_bob) -- node[below] {Verification: accept/reject} (verification_alice);

\end{tikzpicture}
}
\caption{The Fiat-Shamir transform reduces the interactivity of a sigma protocol. Alice uses  $sk_A$ to generate commitments $\comm_i$, $i \in [\kappa]$, and then hashes to generate challenges $\chal_i$, $i \in [\kappa]$. Bob then verifies using Alice's public key $pk_A$, leading to acceptance or rejection and reducing the communication between Alice and Bob. }
\label{fig:fs_simple}
\end{figure}
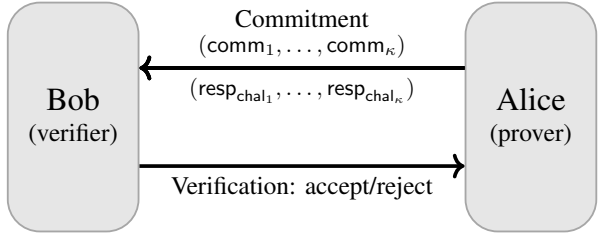

The Fiat-Shamir transform modifies the sigma protocol to construct a digital signature scheme in two key ways:
\begin{description}[leftmargin=0pt]\itemsep0.5em
\item[\textbf{Parallel repetition}:] \footnote{Parallel repetition is not soundness-amplifying for arbitrary interactive arguments, but it {is} for the type used here \cite{AttemaFehr2022}.}
 The prover constructs \(\kappa\) commitments $\comm_1, \dots, \comm_{\kappa}$ for some integer $\kappa$ and sends them to the verifier, who then sends \(\kappa\) challenges $\chal_1, \dots, \chal_{\kappa}$ to the prover. The prover constructs  \(\kappa\) responses $\resp_{\chal_1}, \dots, \resp_{\chal_{\kappa}}$. The verifier accepts the proof (signature) if  each of the \(\kappa\) transcripts \( \{(\comm_i, \chal_i, \resp_i)\}_{i=1}^\kappa\) is accepted.

\item[\textbf{De-interactivization}:] Using a cryptographic hash function \(H\), 
the prover constructs the challenge herself as
\[\resizebox{0.9\columnwidth}{!}{$(\chal_1, \hdots, \chal_\kappa) = H(\comm_1, \hdots, \comm_\kappa, m)$,}\]
where  \(m\) is the message being signed.
\end{description}
If the original protocol has a challenge space of size \(S\),  after parallel repetition, the protocol remains 2-special sound, while the challenge space size increases to \(S^\kappa\), meaning that the \emph{soundness error} is reduced from \(S^{-1}\) to \(S^{-\kappa}\). Thus, it suffices to take \(\kappa = \frac{128}{\log_2c}\) (respectively, \(\frac{192}{\log_2c}, \frac{256}{\log_2c}\) for NIST Level I (respectively, Level III, Level V) security.\footnote{This is in contrast with the classical Schnorr protocol~\cite{JC:Schnorr91}, which does not require parallel repetition because it naturally has exponentially-small soundness error.} De-interactivization then transforms the soundness-amplified sigma protocol into a digital signature scheme as described in~\Cref{fig:fs_simple}.

\rmv{
\begin{figure*}[htbp]
  \centering
  \begin{pchstack}[boxed]
      \procedure[linenumbering, lnstart=100]{$\kgen()$}
      {
	A \sample \Aut(V) \\
	G_1 \gets \rref(G_0A) \\
	\pcreturn (A, G_1)
      }

    \phantom{M}

    \procedure[linenumbering, lnstart=200]{$\sign(A, (G_0, G_1); m)$}
    {
      R_1, \hdots, R_\kappa \sample \Aut(V) \\
      \pcfor i = 1,2, \hdots, \kappa\\
      \pcind \comm_i \gets \rref(G_0R_i) \\
      \vec{\chal} \gets H(\vec{\comm},m) \\
      \pcfor i = 1,2, \hdots, \kappa\\
      \pcind \resp_i \gets A^{-\chal_i}R_i \\
      \sigma = (\vec{\comm},\vec{\resp}) \\
      \pcreturn \sigma
    }
    \phantom{M}

    \procedure[linenumbering, lnstart=300]{$\verify((G_0,G_1); m,\sigma)$}
    {
      \pcparse (\vec{\comm}, \vec{\resp}) \gets \sigma \\
      \vec{\chal} \gets H(\vec{\comm}, m) \\
      \pcfor i=1,2,\hdots, \kappa\\
      \pcind \pcif \rref(G_\chal \resp_i) \neq \comm_i \\
      \pcind[2] \pcreturn \mbox{``Reject''}\\
      \pcreturn \mbox{``Accept''}
    }
  \end{pchstack}
\caption{The digital signature scheme obtained by applying the Fiat-Shamir transform to the sigma protocol in~\Cref{fig:AbstractSigma}.}
\label{fig:GenericDigitalSignature}
\end{figure*}}

In the next two sections, we see how the Fiat-Shamir transform gives rise to two new digital signature schemes, each based on a different perspective.  
In Section \ref{section:CROSS}, 
we consider CROSS, which exemplifies a syndrome-decoding–style approach. Later, in Section \ref{section:LESS}, we consider equivalence-based designs (LESS/MEDS) that trade different forms of structure for different size and performance profiles.

\section{Digital Signature Scheme based on Restricted Syndrome Decoding}
\label{section:CROSS}

Syndrome decoding has been used as a decoding technique for linear codes since the early 1950s, dating back nearly to the beginning of coding theory \cite{Hamming}. In the 1970s, Berlekamp, McEliece, and van Tilborg \cite{BMvT_78} proved that decoding a random linear code in the Hamming metric is NP‑complete. The hardness of syndrome decoding is integral to several code-based KEMs, including McEliece and HQC, though each considers a particular family of codes (binary Goppa for McEliece and binary quasi-cyclic for HQC). CROSS modifies the Syndrome Decoding Problem (SDP), to be stated shortly, and uses it in a novel way to provide a new digital signature scheme. We explore this variant of the SDP and then the signature scheme.

\subsection{Restricted Syndrome Decoding}

As all codes considered in this article are linear, we use the term ``code'' to mean linear code. An $[n,k]$ linear code $C$ over the alphabet $\mathbb{F}_q$, meaning the finite field with $q$ elements, is an $\F_q$-linear subspace of $\mathbb{F}_q^n$.  
Let $\F_q^{m \times n}$ be the set of $m\times n$ matrices with entries in $\mathbb{F}_q$. For an $[n,k]$ code $C$, there exists a parity check matrix $H\in \F_q^{(n-k) \times n}$ with the property that $\x\in C$ if and only if $\x H^T = \mathbf{0}$.  The Hamming weight of a vector $\mathbf{v}=(v_1,\ldots,v_n)$ is the number of nonzero entries, $wt(\mathbf{v})=| \left\{ i : v_i \neq 0 \right\} |$. The syndrome of a vector $\x \in \F_q^n$ (with respect to $H$) is  $\x H^T \in \F_q^{n-k}$.  A code whose minimal nonzero codeword weight is $d$ is said to have minimum distance $d$ and is able to correct any $t$ errors, provided $d \geq 2t+1$. Indeed, given a received word $\w=\mathbf c+\e \in \F_q^n$ with $c \in C$, it follows that $\w$ and $\e$ have the same syndrome. 
\begin{figure}[h!]
\begin{tcolorbox}[colback=yellow!10!white, colframe=green!50!black, title=Syndrome Decoding Problem (SDP)]
    Given a parity-check matrix $H \in \F_q^{(n-k) \times n}$, $\s \in \F_q^{n-k}$, and $t \in [n]$, find $\e \in \F_q^n$ with syndrome
\begin{equation} \label{eq:syndrome_eq}
  \s=\e H^T  
\end{equation}
and weight
\begin{equation} \label{eq:syndrome_wt}
  wt(\e) \leq t. 
\end{equation}
\end{tcolorbox}
\end{figure}

A solution to the Syndrome Decoding Problem allows one to determine the error vector $\e$ and recover the original codeword $\mathbf c$ if $C$ has minimum distance at least $2t+1$. Notice that (\ref{eq:syndrome_eq}) and (\ref{eq:syndrome_wt}) give a linear and a nonlinear constraint, respectively. We see below that (\ref{eq:syndrome_wt}) is replaced in the Restricted Syndrome Decoding Problem (RSDP). 

\rmv{
It is worth mentioning that
Wave~\cite{Wave}, a Round 1 candidate in NIST's Additional Signature Process, signs by solving a decoding problem. This is in contrast with CROSS, which signs by proving knowledge of a decoding solution, capitalizing on the zero-knowledge protocol framework discussed earlier. Consequently, Wave has a slower signing speed than CROSS. There are pros and cons to each approach. Wave has a smaller signature than CROSS but a larger public key. Wave did not advance to the second round of the Additional Signature process.}
\begin{figure}[h!]
\begin{tcolorbox}[colback=yellow!10!white, colframe=green!50!black, title=Restricted Syndrome Decoding Problem  (RSDP)] 
Given a parity-check matrix $H \in \F_q^{(n-k) \times n}$, $\s \in \F_q^{n-k}$, and $\mathcal{E} \subseteq \F_q^n$,  find $\e \in \F_q^n$ with syndrome 
\begin{equation} 
  \s=\e H^T  
\end{equation}
 and 
\begin{equation} \label{eq:r_syndrome_error}
  \e \in \mathcal{E}.
\end{equation}
\end{tcolorbox}
\end{figure}

\subsection{CROSS} 

CROSS is based on the RSDP with a restricted error set $\mathcal{E}$, and improves on the SDP in two main respects. First, RSDP can permit higher-weight errors while still yielding a unique solution, which increases the cost of information-set decoding (ISD) attacks; consequently, smaller RSDP instances may achieve the same security level as larger SDP instances. Second, when componentwise multiplication acts transitively on the restricted vectors, the restriction can also reduce communication.

CROSS is expected to be quantum-secure because it relies on code-based hardness assumptions rather than number-theoretic problems vulnerable to Shor's algorithm. Its \rmv{EUF-CMA} security reduces to the hardness of solving RSDP (and a subgroup variant) over random linear codes \cite{cross_spec}. The underlying Syndrome Decoding Problem is NP-complete \cite{BMvT_78} and underlies long-standing post-quantum schemes such as Classic McEliece. The best known attacks on SDP/RSDP are ISD-style algorithms and refinements, for which known quantum improvements are at most polynomial (e.g., square-root) rather than exponential; accordingly, CROSS parameters are selected to resist both classical and quantum variants. Overall, CROSS assumes average-case hardness of (restricted) syndrome decoding for random linear codes over finite fields.

Notice that if $\mathcal{E} = \mathbb{F}_q^n$, then the RSDP is the classical SDP. We now define a set that plays a crucial role in CROSS. Take
\[
\mathbb{E}:=\left\{ g^i: i \in [z] \right\} \subset \F_p^*,
\]
where $\F_p^*$ represents the multiplicative group, $z$ is a prime that divides $p-1$, and $g \in \F_p^*$ is of order $z$. When we take $\mathcal{E} = \mathbb{E}^n$, the RSDP is NP-complete \cite{CROSS_ISIT}. In this case, the RSDP is expected to have at most one solution if $\log_2(z) \leq (1-Rate)\log_2(p)$,
where $Rate$ represents the rate of the code whose parity check matrix is $H$. Observe that $\mathbb E^n$, under componentwise multiplication, is a commutative group isomorphic to $ \F_z^n$, viewed as \(n\)-tuples of integers modulo $z$ under componentwise addition, via the map 
$\mathbb{E}^n \to \F_z^n$
\[
g=(g^{i_1}, \dots, g^{i_n}) \mapsto \ell(g)=(i_1, \dots, i_n).\]
Such a representation enables CROSS to perform arithmetic faster and to compactly represent an element in $\mathbb{E}^n$.
\rmv{Allow us to emphasize a key point: given any $e, e' \in \mathbb{E}^n$, write 
$e:=g^{i_1}\dots g^{i_n}, e':= g^{i_1'}\dots g^{i_n'}$. Then,
\begin{align*}
&e'=g^{i_1'-i_1}\dots g^{i_n'-i_n}e \text{ and }\\
&\varphi(e')=\varphi(e)(i_1'-i_1, \dots, i_n'-i_n),
\end{align*}
which allows faster arithmetic.} We now take ${G} \leq \mathbb{E}^n$ to be a multiplicative subgroup given by
\[{G} = \left<a_1,\ldots,a_m\right> = \left\{\prod_{i=1}^m a_i^{u_i} : u_i \in [z] \right\},\]
for some $a_1,\ldots, a_m \in \mathbb{E}^n$ and $m < n$. The RSDP(${G}$), which is still NP-hard, is defined as the RSDP with $\mathcal{E} = {G}$. The RSDP($G$) is expected to have a unique solution provided $|G|p^{-(1-R)n} \leq 1$.

The CROSS signature scheme is obtained after applying the Fiat-Shamir transform on $t$ parallel executions of the following zero-knowledge protocol.
\begin{description}[leftmargin=0pt, itemsep=3pt]
\item[\textbf{Public parameters:}] $G \leq \mathbb{E}^n$
\item[\textbf{Public keys:}] $H \in \F_p^{(n-k) \times n}$, and $\s=\e H^T$.
\item[\textbf{Private key:}] $\mathbf{e} \in G$.
\item[\textbf{Commitment phase:}] The prover samples \(\mathbf{Seed}\) uniformly at random from \(\{0,1\}^\lambda\). Then, through a deterministic cryptographically secure pseudorandom generator, she makes the following samples: \( \left( \mathbf{Seed}^{(\mathbf{u}')},\mathbf{Seed}^{(\mathbf{e}')}\right)\) from \(\{0,1\}^{2\lambda}\) with initialization \(\mathbf{Seed}\), \(\mathbf{u}'\) from \(\F_p^n\) with initialization \(\mathbf{Seed}^{(\mathbf{u}')}\), and  \(\e'\) from \(G\) with initialization \(\mathbf{Seed}^{(\mathbf{e}')}\). She computes $\sigma \in G$ such that\footnote{As the element $\sigma \in G$ gives rise to a bijection $G \to G$, we denote the product $\sigma \e'$ as $\sigma(\e')$ because the same bijection is used later.} \(\sigma(\e')=\e\). Define  \(\mathbf{u}=\sigma(\mathbf{u}')\) and compute $\Tilde{\s}= \mathbf{u} H^T$. She constructs the \emph{commitment} as \[ (c_0,c_1) = (\hash(\Tilde{\s},\sigma),\hash(\mathbf{u}',\e')),\] and sends it to the verifier.
\item[\textbf{Challenge phase 1:}] The verifier samples a \emph{challenge} \(\beta\) uniformly at random from \( \F_p^* \), and sends it to the prover.
\item[\textbf{Response phase 1:}] The prover computes $\mathbf{y}=\mathbf{u}'+\beta\e'$ and sends \( h = \hash(\mathbf{y})\) to the verifier.
\item[\textbf{Challenge phase 2:}] The verifier samples a \emph{challenge} \(b\) uniformly at random from \( \{0,1\} \), and sends it to the prover.
\item[\textbf{Response phase 2:}] The prover constructs
\[
f = \begin{cases}
\left(\mathbf{y},\sigma\right) & \mbox{if } b = 0 \\
\mathbf{Seed} & \mbox{if } b = 1
\end{cases}
\]
and sends it to the verifier.
\item[\textbf{Verification:}] The verifier checks the following.
If $b=0$, he computes $\Tilde{\mathbf{y}}=\sigma(\mathbf{y})$ and $\Tilde{\s} = \Tilde{\mathbf{y}} H^T - \beta \s$.

Thus,
$\Tilde{\mathbf{y}} = \sigma(\mathbf{y}) = \sigma(\mathbf{u}^\prime) + \sigma(\beta\e^\prime) = \mathbf{u} + \beta \e$ and
\[
\Tilde{\s} = \Tilde{\mathbf{y}} H^T - \beta \s
= \mathbf{u}H^T + \beta \e H^T - \beta \s = \mathbf{u}H^T.\]
He accepts if $\hash(\mathbf{y})=h$, $\hash(\Tilde{\s},\sigma) = c_0$, and $\sigma \in G$. If $b=1$, he uses the deterministic cryptographically secure pseudorandom generator to sample \( \left( \mathbf{Seed}^{(\mathbf{u}')},\mathbf{Seed}^{(\mathbf{e}')}\right)\) from \(\{0,1\}^{2\lambda}\) with initialization \(\mathbf{Seed}\). Compute $\mathbf{y} = \mathbf{u}' + \beta \mathbf{e}'$. He accepts if $\hash(\mathbf{y})=h$ and $\hash(\mathbf{u}',\mathbf{e}') = c_1$.
\end{description}
This zero-knowledge protocol is a type of $5$-pass identification scheme called a $q2$-Identification scheme, with $q=p-1$, because the first challenge can take $q$ different values, while the second can take only 2 (0 or 1). In particular, CROSS is a variant of the $5$-pass identification scheme first introduced by \cite{q2IDS}, which implies that CROSS achieves a level of security known as EUF-CMA.

\section{Digital Signature Schemes based on Code Equivalence}
\label{section:LESS}

\subsection{The Code Equivalence Problem}

It is natural to ask when two mathematical objects are essentially the same. Perhaps the earliest reference to this notion for codes is Golay's work~\cite{Golay1949Notes}, in which codes are considered up to coordinate relabeling. The concept was then formalized in MacWilliams' dissertation~\cite{MacWilliams_dissertation} to the notion we now call code equivalence, to be defined shortly. Code equivalence made its appearance in cryptography in the late 1970s with the work of Berlekamp, McEliece, and van Tilborg \cite{BMvT_78}. The code equivalence decision problem asks if two codes are the same up to a weight-preserving  isomorphism (for instance, a coordinate permutation). It was shown in the late 1990s that the graph isomorphism problem can be reduced to the code equivalence problem in polynomial time \cite{PetrankRoth1997}. Babai demonstrated a quasi-polynomial-time algorithm for the graph isomorphism problem \cite{Babai2016GI}, but it is still unknown whether either problem admits a polynomial time algorithm. 

When considering code equivalence between two codes, a necessary condition is that both codes have the same parameters or the same weight enumerators. However, it is not always easy to determine these values. For instance, for the small example in~Figure \ref{fig:EquivalentTernaryCodes}, it is feasible to see that the two codes are identical up to interchanging two coordinates because the codes are short (length $3$) with only $4$ codewords. As we see below, the inability to quickly identify large codes that are the same up to relabeling (for instance) underpins some digital signature schemes.

Let \(C_0, C_1 \subseteq V\) be two codes in the same ambient space \(V\) over the field \(\FF_q\) with weight function \(wt\)---we take \(V = \FF_q^n\) with the Hamming weight and \(V = \F^{n\times m}\) with the rank weight for LESS and MEDS, respectively; this formalism applies to any weighted vector space over \(\FF_q\). An \emph{isomorphism} from \(C_0\) to \(C_1\) is a function \(\psi\colon V \to V\) that is:
\begin{itemize}[wide]
    \item 
linear, i.e. \(\forall a,b \in \FF_q, \x, \y \in V\) we have
\[
\psi(a \x + b \y) = a\psi(\x) + b\psi(\y),
\]
\item weight-preserving, i.e.,
\[
wt(\psi(\x)) = wt(\x) ~\forall \x\in V, \quad \text{and}
\]
\item maps \(C_0\) to \(C_1\), i.e.,
\[
\psi(C_0) = \{ \psi(\x) \,:\, \x \in C_0\} = C_1.
\]
\end{itemize}

When there is an isomorphism from \(C_0\) to \(C_1\), we say that \(C_0\) and \(C_1\) are \emph{isomorphic} (or \emph{equivalent}). 

\begin{tcolorbox}[colback=yellow!10!white, colframe=green!50!black, title=Computational Code Equivalence Problem]
Given two equivalent codes \(C_0\) and \(C_1\), find an isomorphism \(\psi\) from \(C_0\) to \(C_1\). 
\end{tcolorbox}

Computational code equivalence is the hard problem that underlies both LESS and MEDS. The difficulty of the problem depends heavily on the parameters and features of the codes being considered, as well as on the representation of the codes \(C_0\) and \(C_1\). The standard approach is to use the reduced row echelon form (RREF). In particular, if \(C_0 \subseteq \FF_q^n\) is a code of dimension \(k\), then \(C_0\) can be written as the row span of a \(k \times n\) generator matrix \(G_0\).\footnote{While this discussion applies to LESS directly, some extra work is required for MEDS---see~\Cref{sec:MEDS}.} Two generator matrices \(G_0, G_0'\) will generate the same code \(C_0\) if and only if they are row equivalent, and it follows that \(C_0\) can be represented by a unique RREF generator matrix. For a given matrix \(G\), we denote by \(\rref(G)\) its reduced row echelon form.

For any linear map \(\psi \colon V \to V\), there exists a \emph{matrix representation} \(A_\psi\) of \(\psi\) with respect to the standard basis. In particular, \(A_\psi\) is defined by
$
\psi(\x) = \x A_\psi~\forall \x \in V$. If \(G_0\) is the RREF generator matrix for \(C_0\), then \(\rref(G_0A_\psi)\) is the RREF generator matrix for \(C_1 = \psi(C_0)\). This gives a convenient way to formulate code equivalence in terms of RREF generator matrices: the codes generated by \(G_0\) and \(G_1\) are equivalent if and only if there is a matrix \(A\)  such that \(G_1 = \rref(G_0A)\) and which is weight-preserving on the ambient space.
\rmv{; i.e.,
\begin{equation}
\label{eqn:WeightPreserving}
wt(\x A) = wt(\x)~\forall 
\x \in V.
\end{equation}}
We denote by \(\Aut(V)\) the set of all such matrices, called the \emph{automorphism group} of \( (V, wt)\).

\subsection{Digital Signatures from Code Equivalence}
\label{sec:DSfromCE}

At a high level, LESS and MEDS are built from a sigma protocol in the following framework. 
\begin{itemize}[leftmargin=0pt, label={}]
\item \textbf{Public parameters:} All users agree on an ambient space \(V\) with weight function \(wt\).
\item \textbf{Public and private keys:} The prover's public key is a pair of RREF generator matrices \( (G_0,G_1)\) with \(G_1 \in \lang_{G_0}\). The prover's secret key is \(A \in \Aut(V)\) such that \(G_1 = \rref(G_0A)\). 
\item \textbf{Commitment phase:} The prover samples \(R\) uniformly at random from \(\Aut(V)\), and constructs the \emph{commitment} as
$\comm = \rref(G_0R)$.
She sends \(\comm\) to the verifier.
\item \textbf{Challenge phase:} The verifier samples a \emph{challenge} \(\chal\) uniformly at random from \(\{0,1\}\), and sends it to the prover.
\item \textbf{Response phase:} The prover constructs
\[
\resp
=
A^{-\chal}R
= \begin{cases}
R & \mbox{if } \chal = 0 \\ A^{-1}R & \mbox{if } \chal = 1
\end{cases}
\]
and sends it to the verifier.
\item \textbf{Verification:} The verifier checks that
\[
\begin{cases}
\rref(G_0\resp) = \comm & \mbox{if } \chal = 0 \\
\rref(G_1\resp) = \comm & \mbox{if } \chal = 1
\end{cases}
\]
which we write more compactly as \(\rref(G_\chal\resp) = \comm\).
\end{itemize}
\rmv{
\begin{figure}[htb]
\centering
\scriptsize
  \fbox{
  \tabcolsep4pt
    \begin{tabular}{lcl}
      \(\prover: ( (G_0, G_1), A) \) && \(\verifier: (G_0, G_1)\) \\\hline
      \(R \sample \Aut(V)\) \\
      \(\comm \gets \rref(G_0R)\) & \rsends[0pt]{\comm} \\
      &\lsends[0pt]{\chal}& \(\chal \sample \{0,1\}\) \\[6pt]
      \(\resp \gets A^{-\chal}R\) & \rsends[0pt]{\resp} & \(\begin{array}{l} \mbox{Accept if } \\ \rref(G_\chal\resp) = \comm \end{array}\)
    \end{tabular}
  }
\caption{The high-level structure of the sigma protocol used in both LESS and MEDS.}
\label{fig:AbstractSigma}
\end{figure}}
We call a triple \(\tau = (\comm, \chal, \resp)\) a \emph{transcript}, and we say that a transcript is \emph{accepting} if it would result in \(\verifier\) accepting the proof; equivalently, if
\[
\rref(G_\chal \resp) = \comm.
\]
The public information consists of \(G_0 \) and \(G_1\) while the secret key is \(A\); thus, for security, we require that it must be difficult to recover any \(A'\) such that \(G_1 = \rref(G_0A')\) given \(G_0\) and \(G_1\). To transform this protocol into a correct and secure digital signature scheme, it must satisfy three correctness and security conditions:
\begin{itemize}[leftmargin=0pt, itemsep=3pt]
\item[] \textbf{Perfect Correctness}: If the protocol is executed honestly, then \(\verifier\) will accept the proof.
\item[] \textbf{2-Special Soundness}: There must be an efficient algorithm \(\Ext\)---the \emph{extractor}---which, on input two accepting transcripts
$$
\tau_0 = (\comm, 0, \resp_0), \mbox{ and }
\tau_1 = (\comm, 1, \resp_1)
$$
\rmv{\begin{align*}
\tau_0 &= (\comm, 0, \resp_0) \\
\tau_1 &= (\comm, 1, \resp_1)
\end{align*}}
with the same commitment, outputs a witness \(A' = \Ext(\tau_0, \tau_1)\) such that \(G_1 = \rref(G_0A')\).
\item[] \textbf{Honest Verifier Zero-Knowledge}: There must exist an efficient algorithm \(\Sim\) that, given \(G_0\) and \(G_1\) (but \emph{not} \(A\)) produces an accepting transcript
\(
\tau = (\comm, \chal, \resp).
\)
Moreover, the distribution of transcripts obtained from \(\Sim\) must be identical to the distribution of transcripts output by honest interactions between \(\prover\) and \(\verifier\).
\end{itemize} 
Intuitively, 2-special soundness ensures that any prover convincing the honest verifier with probability greater than $\frac{1}{2}$ knows a corresponding witness (secret key). Honest-verifier zero-knowledge ensures the verifier learns nothing about that secret  when interacting with a knowledgeable prover. Both properties are verified directly; see, for example, LESS~\cite[Section 4]{AFRICACRYPT:BMPS20} and MEDS~\cite[Section 4]{AFRICACRYPT:CNPRRS23}.

\subsection{LESS: Linear Equivalence Signature Scheme} 

In this section, we describe LESS, a signature scheme based on (linear) code equivalence, and we occasionally reference MEDS, a closely related construction based on matrix code equivalence, for brief points of comparison. A detailed treatment of MEDS is deferred to Section~\ref{sec:MEDS}.

LESS takes \(V = \FF_q^n\) with the Hamming weight \(wt(\x) = |\{i \in [n] \,:\, x_i \neq 0\}|\), and considers codes \(C \subseteq \FF_q^n\) of dimension \(k\). 
\begin{figure}[h!]
\centering
\includegraphics[width=0.75\columnwidth]{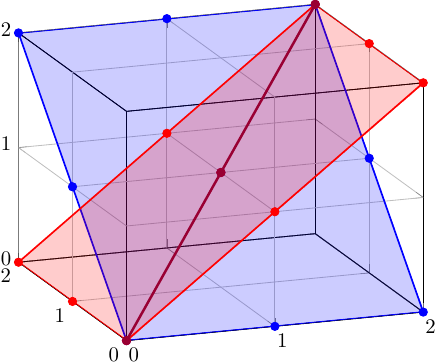}
\caption{Two equivalent linear codes over \(\FF_3\). In blue, the code \(C_0\) is generated by \( (1,0,0)\) and \( (0,1,1)\). In red, the code \(C_1\) is generated by \( (0,1,0)\) and \((1,0,1)\). These codes are equivalent: \(C_1\) is obtained from \(C_0\) by exchanging the first and second coordinates. Points in purple are common to both codes.}
\label{fig:EquivalentTernaryCodes}
\end{figure}

Two kinds of linear maps that preserve this weight are
\begin{itemize}[leftmargin=0pt, itemsep=3pt, label={}]
\item \textbf{Scaling coordinates:} For \(\boldsymbol{\lambda} = (\lambda_1, \hdots, \lambda_n), \x=(x_1,\dots,x_n) \in \FF_q^n\), define the \emph{Hadamard product} 
\(
\boldsymbol{\lambda} \circ \x = (\lambda_1x_1, \hdots, \lambda_nx_n).
\)
If \(\boldsymbol{\lambda} \in (\FF_q^*)^n\), then
\(
\wt(\boldsymbol{\lambda}\circ \x) = \wt(\x)
\) for any \(\x\).
\item \textbf{Permuting coordinates:} For any permutation \(\sigma \) in the symmetric group \( S_n\), define
\(
\sigma \ast \x = (x_{\sigma^{-1}(1)}, x_{\sigma^{-1}(2)}, \hdots, x_{\sigma^{-1}(n)}).
\)
Then, \(\wt(\sigma \ast \x) = \wt(\x)\) for all \(\x \in \FF_q^n\).
\end{itemize}
According to the MacWilliams Theorem on Equivalence of Codes~\cite[Theorem 1]{macwilliams}, these maps are essentially the only weight-preserving linear transformations of \(\FF_q^n\) with the Hamming metric. Thus, any such map \(\psi\) can be written as
\(
\psi \colon \x \to \sigma \ast (\boldsymbol{\lambda} \circ \x)
\)
for some \(\boldsymbol{\lambda} \in (\FF_q^*)^n\) and some \(\sigma \in S_n\). Such a map \(\psi\) has matrix representation
\(
A_\psi = DP
\)
where \(D\) is the diagonal matrix with \(D_{ii} = \lambda_i~\forall i\in [n]\), and \(P\) is the permutation matrix of \(\sigma\):
\[
P = \sum_{i=1}^n \e_{\sigma(i)}\e_i^T.
\]
A matrix of the form \(A = DP\) is called a \emph{monomial matrix}, and the collection of \(n\times n\) monomial matrices over \(\FF_q\) is denoted \(\Mon(\FF_q)\). When \(wt\) is the Hamming weight, \(\Aut(\FF_q^n, wt) = \Mon(\FF_q)\). \rmv{Figure~\ref{fig:EquivalentTernaryCodes} depicts two equivalent codes in \(\FF_3^3\), equipped with the Hamming weight. }

\rmv{
In LESS and MEDS, the language and corresponding relation take the form
\begin{align*}
\lang_{G_0} &{=}~\{ G_1 \,:\, \exists A \in \Aut(V) \mbox{ s.t. } G_1 = \rref(G_0A)\}\\
\relation_{G_0} &{=}~\{ (G_1, A) \,:\, A \in \Aut(V), G_1 = \rref(G_0A)\}
\end{align*}
for some ambient space \(V\) with weight \(wt\), and some \(k\times n\) generator matrix \(G_0\).
}

\subsection{MEDS: Matrix Equivalence Digital Signature} 
\label{sec:MEDS}

In MEDS, the ambient space is \(V=\F^{n\times m}\), and the weight function is the rank: \(\wt(X) = \rank(X)\). A linear map \(\psi \colon \F^{n\times m} \to \F^{n\times m}\) can be written as
\[
\psi(X) = A_1XB_1 + A_2XB_2 + \cdots + A_t X B_t
\]
for some \(t \leq (nm)^2\). Naturally, these maps will generally not be weight-preserving on \(\F^{n\times m}\). It is known that the only such maps that preserve rank take the form
\(
\psi(X) = AXB, \mbox{ or } \psi(X) = AX^TB
\)
for some \(A \in \GL_n(\FF_q)\) and \(B \in \GL_m(\FF_q)\), and moreover that the second case is possible only when \(n = m\)~\cite[Theorem 1.3]{rankmacwilliams}. Following the MEDS specification, we will consider only maps of the first kind. Thus, the isomorphisms of codes take the form
\(
C_0 \mapsto C_1 = AC_0B
\)
for invertible \(A\) and \(B\) of the correct size.

To make this formulation of MEDS compatible with our generic description in Section~\ref{sec:DSfromCE}, we need a generator matrix formulation of code equivalence for matrix codes \(C_0 \subseteq \F^{n\times m}\). The standard approach is to \emph{vectorize} the code: for \(X \in \F^{n\times m}\), we define \(\Vec(X)\) to be the vector obtained from \(X\) by concatenating its rows:
$ 
\Vec(X) = [\x_1,  \hdots, \x_n]$.
The vectorization of a code \(C_0 \subseteq \F^{n\times m}\) is
\[
\Vec(C_0) = \{\Vec(X) : X \in C_0\} \subseteq\FF_q^{nm}.
\]
If \( \{X_1, \hdots, X_k\}\) is a basis for \(C_0\), then \( \{\Vec(X_1), \hdots, \Vec(X_k)\}\) is a basis for \(\Vec(C_0)\), which we can arrange as the rows of a matrix and row reduce, yielding an RREF generator matrix \(G_0\) for \(\Vec(C)\). This vectorization interacts nicely with code equivalence.  We have the identity
\[
\Vec(AXB) = \Vec(X) (A^T \otimes B)
\]
where \(\otimes\) is the \emph{Kronecker product}, defined by
\[
A \otimes B = \begin{bmatrix} 
a_{11}B & a_{12}B & \cdots & a_{1\ell}B \\ 
\vdots  &  \vdots & \ddots & \vdots    \\
a_{k1}B & a_{k2}B & \cdots & a_{k\ell}B 
\end{bmatrix}
\]
when \(A = (a_{ij})_{\substack{1 \leq i \leq k \\ 1 \leq j \leq \ell}}\). Hence, if \(G_0\) is the RREF generator matrix for \(C_0\), then
\(
G_1 = \rref(G_0 (A^T\otimes B))
\)
is the RREF generator matrix for \(\Vec(AC_0B)\). 

\subsection{Implementation Details}

LESS and MEDS feature optimizations to improve signing time and signature size. In this section, we detail a number of these optimizations.

\subsubsection{Multiple Public Keys}
\begin{figure}[htb]
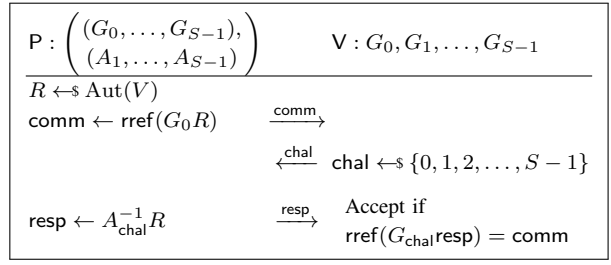

\centering
\scriptsize
  \fbox{
  \tabcolsep1pt
    \begin{tabular}{lcl}
      {\(\prover: \left(\arraycolsep1pt\begin{array}{c} (G_0,\hdots, G_{S-1}), \\ (A_1, \hdots, A_{S-1})\end{array}\right) \)} && \(\verifier: G_0, G_1, \hdots, G_{S-1}\) \\\hline
      \(R \sample \Aut(V)\) \\
      \(\comm \gets \rref(G_0R)\) & \rsends[0pt]{\comm} \\
      &\lsends[0pt]{\chal}& \(\chal \sample \{0,1,2,\hdots,S-1\}\) \\[6pt]
      \(\resp \gets A^{-1}_{\chal}R\) & \rsends[0pt]{\resp} & \(\begin{array}{l} \mbox{Accept if } \\ \rref(G_\chal\resp) = \comm \end{array}\)
    \end{tabular}
  }
\caption{The sigma protocol incorporating multiple public keys. Here, \(A_0 = I\), the identity matrix.}
\label{fig:MPK}
\end{figure}

To reduce the number of repetitions required, we can increase the challenge space size by increasing the number of codes in the public key. We fix an integer \(S \geq 2\) and define the secret key as a tuple \( (A_1, \hdots, A_{S-1}) \sample \Aut(V)^{S-1}\) with corresponding public key \( (G_0, G_1, \hdots, G_{S-1})\), where \(G_i = \rref(G_0A_i)\). Then, we adapt the sigma protocol as depicted in~\Cref{fig:MPK}.

Since the protocol of~\Cref{fig:MPK} now has \(S\) possible challenges, its soundness error falls to \(S^{-1}\). This means that, in order to achieve soundness error \(2^{-\lambda}\), it suffices to take \(\kappa = \frac{\lambda}{\log_2 S}\) parallel repetitions of the protocol. This optimization decreases signature size and running time, at the cost of increasing public key sizes. Both LESS and MEDS consider \(S = 2\) (corresponding to a single public key), while LESS also proposes parameter sets with \(S = 4, 8\).

\subsubsection{Exploiting Commitment Recoverability}

\begin{figure}[htb!]
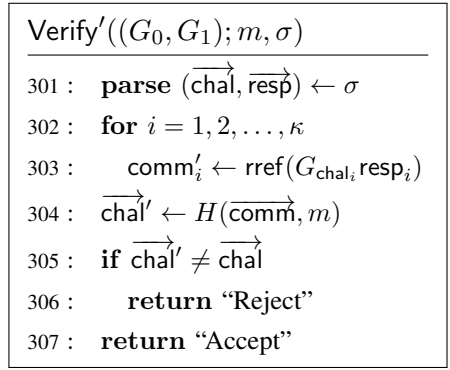

\centering
\fbox{
    \procedure[linenumbering, lnstart=300]{$\verify'( (G_0,G_1); m,\sigma)$}
    {
      \vphantom{\mbox{\Huge I}}\pcparse (\vec{\chal}, \vec{\resp}) \gets \sigma \\
      \pcfor i = 1,2,\hdots,\kappa \\
      \pcind \comm'_i \gets \rref(G_{\chal_i}\resp_i)\\
      \vec{\chal}' \gets H(\vec{\comm}, m) \\
      \pcif \vec{\chal}' \neq \vec{\chal}\\
      \pcind \pcreturn \mbox{``Reject''}\\
      \pcreturn \mbox{``Accept''}
    }
    }
    \caption{The modified verification procedure exploiting commitment recoverability.}
    \label{fig:CRVer}
\end{figure}

\emph{Commitment recoverable} schemes are such that given a public key \( (G_0, G_1)\) and a challenge/response pair \( (\chal, \resp)\), there is an efficient algorithm that recovers the unique accepting commitment \(\comm\). In LESS and MEDS, we must have
$
\comm = \rref(G_\chal\resp).
$
This enables a straightforward optimization of the protocol: the signature will instead consist of the \emph{challenge} string and response vector: \(\sigma = (\vec{\chal}, \vec{\resp})\). During verification, the verifier will reconstruct the commitment vector using the signature, recompute the challenge by hashing the reconstructed commitment vector and message, and accept the proof if and only if the hash output is equal to the challenge component of the signature as depicted in Figure~\ref{fig:CRVer}.
The commitment vector consists of \(\kappa\) elements of \(\Aut(V)\)---requiring approximately \(\kappa \cdot (n\log_2 q + \log_2 n!)\) bits for LESS, and \(\kappa \cdot (n^2+m^2)\log_2 q\) bits for MEDS---while the challenge vector consists of \(\kappa\) integers between \(1\) and \(S\), which can be represented in approximately \(\kappa\log_2 S\) bits. For proposed LESS and MEDS parameter sets, this optimization yields a substantial decrease in signature size.

\subsubsection{Random Seeds and Constant-Weight Hash Functions}

When \(\resp\) is an element of \(\Aut(V)\), it requires approximately \(n\log_2 q + \log_2 n!\) (for LESS) or \( (n^2+m^2)\log_2 q\) (for MEDS) bits to represent. \rmv{However,  when the \(\chal = 0\), \(\resp\) is simply the ephemeral secret that is used when constructing \(\comm\), which in practice is generated from a \(\lambda\)-bit seed when targeting \(\lambda\)-bit security, using a fixed, public pseudorandom generator.} A straightforward optimization of the protocol is to send the random seeds for the rounds when \(\chal = 0\), rather than the group element \(R\).

To keep responses—and thus the final signature—small, it is preferable that the challenge string have low Hamming weight. This can be enforced via a \emph{constant-weight hash function} \(H\), whose outputs are binary strings of length \(\kappa\) with exactly \(w\) nonzero entries. It suffices to choose \(\kappa\) and \(w\) with
\[
\binom{\kappa}{w}(S-1)^w \ge 2^\lambda.
\]
Compared to an ordinary hash with codomain \(\{0,1\}^\kappa\), achieving the same security typically requires larger \(\kappa\); in practice, the resulting increase in rounds is outweighed by the signature-size reduction from having more rounds where the response is a seed rather than an element of \(\Aut(V)\).

\section{Performance Comparison}

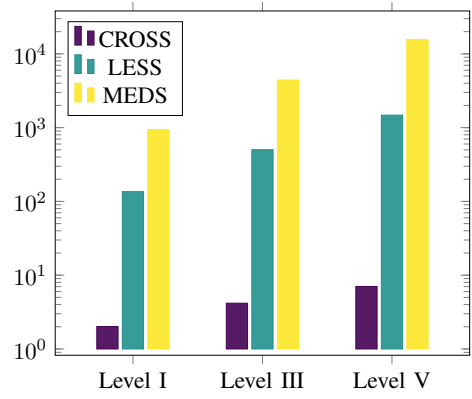
\begin{figure}[h!]
\centering
\begin{tikzpicture}[scale=0.8]
    \begin{semilogyaxis}[
            ybar,
            symbolic x coords={Level I, Level III, Level V},
            xtick=data,
            legend pos = north west,
            enlarge x limits=0.3,
            colormap/viridis
        ]
        \addplot[viridis1, fill=viridis1!90] table[x=Security Level,y=CROSS]{\signtime};
        \addplot[viridis3, fill=viridis3!90] table[x=Security Level,y=LESS]{\signtime};
        \addplot[viridis5, fill=viridis5!90] table[x=Security Level,y=MEDS]{\signtime};
        \legend{CROSS,LESS,MEDS}
    \end{semilogyaxis}
\end{tikzpicture}
\caption{Signing time in Mcycles.}
\label{fig:SigTime}
\end{figure}

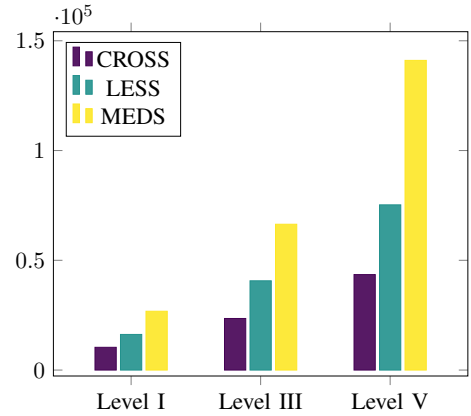
\begin{figure}[h!]
\centering
\begin{tikzpicture}[scale=0.8]
    \begin{axis}[
            ybar,
            symbolic x coords={Level I,Level III,Level V},
            xtick=data,
            legend pos = north west,
            enlarge x limits=0.3,
        ]
        \addplot[viridis1, fill=viridis1!90] table[x=Security Level,y=CROSS]{\pksigsize};
        \addplot[viridis3, fill=viridis3!90] table[x=Security Level,y=LESS]{\pksigsize};
        \addplot[viridis5, fill=viridis5!90] table[x=Security Level,y=MEDS]{\pksigsize};
        \legend{CROSS,LESS,MEDS}
    \end{axis}
\end{tikzpicture}
\caption{Public key + signature size in bytes.}
\vspace{-12pt}
\label{fig:Size}
\end{figure}

Figures~\ref{fig:SigTime} and~\ref{fig:Size} compare the signing time and (public key + signature) sizes for CROSS, LESS, and MEDS at NIST security levels I, III, and V.\footnote{New MEDS parameters taken from \url{https://groups.google.com/a/list.nist.gov/g/pqc-forum/c/pbT_DnPrc2A/m/ZPrIVSmFCQAJ}}

\pgfplotstableread[row sep=\\,col sep=&]{
    Security Level  & CROSS & LESS   & MEDS    \\
    Level I         & 2.013 & 135.6  & 938.6   \\
    Level III       & 4.171 & 502.7  & 4415.6  \\
    Level V         & 7.042 & 1476.6 & 15585.7 \\
    }\signtime
    
\pgfplotstableread[row sep=\\,col sep=&]{
    Security Level  & CROSS & LESS   & MEDS  \\
    Level I         & 10365 & 16229  & 26795 \\
    Level III       & 23498 & 40659  & 66426 \\
    Level V         & 43494 & 75257  & 141068 \\
    }\pksigsize

\section{Conclusion} 
\label{section:conclusion}
This article provided an overview of modern code-based digital signatures based on restricted syndrome decoding and code equivalence: CROSS, LESS, and MEDS. 
These protocols are believed to be quantum-safe. Although all are code-based signatures, they embody distinct design philosophies and exhibit different performance. CROSS is built around a restricted form of syndrome decoding, whereas LESS and MEDS are based on code equivalence problems.

All three code-based digital signatures are based on novel underlying assumptions and feature a variety of performant parameter sets for varied use cases. The concrete security of the underlying problems remains an important direction for future research.

\bibliography{abbrev3,bib,crypto}{}
\bibliographystyle{abbrv}

\newpage

\section*{Short Bios}

\noindent
Sarah Arpin (sarpin@vt.edu) is an Assistant Professor in the Department of Mathematics at Virginia Tech. She earned an M.A. in Pure Mathematics from CUNY Hunter College, an M.S. in Applied Mathematics, and a Ph.D. in Number Theory from the University of Colorado Boulder. She completed a postdoc jointly with Leiden University and the Quantum Software Consortium in the Netherlands. Her research interests include number theory, cryptography, and coding theory.\\[-1pt]

\noindent
Jason LeGrow (jlegrow@vt.edu) is an Assistant Professor in Virginia Tech’s Mathematics Department. Previously, he was a research fellow at the University of Auckland, Mathematics Department. He holds a Ph.D. and MMath in combinatorics and optimization, both from the University of Waterloo, and a BSc (Hons) in pure mathematics from Memorial University of Newfoundland. His research interests are in post-quantum cryptography.\\[-1pt] 

\noindent
Hiram H. López (hhlopez@vt.edu) is an Associate Professor in the Department of Mathematics at Virginia Tech. He received the B.S. degree in applied mathematics from the Autonomous University of Aguascalientes and the Ph.D. in mathematics from CINVESTAV-IPN. He held a postdoctoral position at Clemson University and a tenure-track position at Cleveland State University. His research interests include coding theory, commutative algebra, and image processing.\\[-1pt]

\noindent
Gretchen Matthews (gmatthews@vt.edu) is a Professor of Mathematics at Virginia Tech and Director of a regional component of the Commonwealth Cyber Initiative (CCI). Matthews earned a B.S. from Oklahoma State University and a Ph.D. from Louisiana State University, both in mathematics, and an M.B.A. from Virginia Tech. She held a postdoctoral appointment at the University of Tennessee and was on the faculty at Clemson University. Her research interests include algebraic geometry and combinatorics and their applications to coding theory and cryptography.\\

\rmv{\noindent
Gretchen Matthews is a Professor of Mathematics at Virginia Tech and Director of a regional component of the Commonwealth Cyber Initiative (CCI). Matthews earned a B.S. from Oklahoma State University, a Ph.D. from Louisiana State University (both in mathematics), and an M.B.A. from Virginia Tech. She held a postdoctoral appointment at the University of Tennessee and was on the faculty at Clemson University. Her research interests include algebraic geometry and combinatorics and their applications to coding theory and cryptography.\\

\noindent
Hiram H. L\'opez is an Assistant Professor in the Department of Mathematics at Virginia Tech. He held positions as an Assistant Professor at Cleveland State University and as a Postdoctoral Fellow at Clemson University. He received a Ph.D. in mathematics from CINVESTAV-IPN and a B.S. in applied mathematics from the Autonomous University of Aguascalientes. His research interests include coding theory, commutative algebra, and image processing.\\}

\end{document}